\documentclass[prb,twocolumn,showpacs,preprintnumbers,amsmath,amssymb]{revtex4}
\usepackage{graphicx}% Include figure files
\usepackage{dcolumn}% Align table columns on decimal point
\begin{document}

% Use the \preprint command to place your local institutional report
% number in the upper righthand corner of the title page in preprint mode.
% Multiple \preprint commands are allowed.
% Use the 'preprintnumbers' class option to override journal defaults
% to display numbers if necessary
%\preprint{}

%Title of paper
\title{ Evolution of superconductivity in Pr$_{1-x}$La$_{x}$Os$_{4}$Sb$_{12}$: 
Upper critical field measurements }

% repeat the \author .. \affiliation  etc. as needed
% \email, \thanks, \homepage, \altaffiliation all apply to the current
% author. Explanatory text should go in the []'s, actual e-mail
% address or url should go in the {}'s for \email and \homepage.
% Please use the appropriate macro foreach each type of information

% \affiliation command applies to all authors since the last
% \affiliation command. The \affiliation command should follow the
% other information
% \affiliation can be followed by \email, \homepage, \thanks as well.
\author{C.R. Rotundu}
\author{H. Tsujii}%
\altaffiliation{Present address: Department of Physics, Kanazawa University, Kakuma-machi, 
Kanazawa 920-1192, Japan}
\author{P. Kumar}
\author{B. Andraka}%
\email[Author to whom correspondence should be
addressed:]{andraka@phys.ufl.edu}
\affiliation{ Department of Physics, University of Florida\\
P.O. Box 118440, Gainesville, Florida  32611-8440, USA }
\author{H. Sugawara}
\affiliation{ Department of Mathematical and Natural Sciences,
University of Tokushima\\ Minami-jyousanjima-machi 1-1, Tokushima 770-8502, 
Japan} 
\author{Y. Aoki and H. Sato}
\affiliation{ Department of Physics, Tokyo Metropolitan University\\
Minami-Ohsawa, 1-1, Hachioji, Tokyo, 192-0397, Japan}
%\homepage[]{}
%\thanks{}
%\altaffiliation{}

%Collaboration name if desired (requires use of superscriptaddress
%option in \documentclass). \noaffiliation is required (may also be
%used with the \author command).
%\collaboration can be followed by \email, \homepage, \thanks as well.
%\collaboration{}
%\noaffiliation

\date{\today}

\begin{abstract}
Measurements of the upper critical field $H_{c2}$ near $T_c$ of Pr$_{1-
x}$La$_{x}$Os$_{4}$Sb$_{12}$ were performed by specific heat. A positive 
curvature in $H_{c2}$ versus $T$ was observed in samples and concentrations 
exhibiting two superconducting transitions. These results argue against this 
curvature being due to two-band superconductivity. The critical field slope -
$dH_{c2}/dT$ suggests the existence of a crossover concentration $x_{cr} \approx 
0.25$, below which there is a rapid suppression of effective electron mass with 
La-alloying. This crossover concentration was previously detected in the 
measurement of the discontinuity of $C/T$ at $T_c$.  

\end{abstract}
% insert suggested PACS numbers in braces on next line
%\pacs{ }
% insert suggested keywords - APS authors don't need to do this
%\keywords{}

%\maketitle must follow title, authors, abstract, \pacs, and \keywords
\maketitle

% body of paper here - Use proper section commands
% References should be done using the \cite, \ref, and \label commands
\section{Introduction}
% Put \label in argument of \section for cross-referencing
%\section{\label{}}
%$\subsection{}
%$\subsubsection{}
PrOs$_{4}$Sb$_{12}$, a new heavy fermion superconductor\cite{Bauer},
remains under extensive research scrutiny for a number of
reasons. First of all, it clearly demonstrates that the heavy
fermion state can exist in a system based on non-Kramers
$f$-ions, such as Pr. This opens a possibility for novel
electronic states and novel microscopic mechanisms leading to the
heavy fermion behavior, which may also be applicable to many
previously investigated systems, particularly those based on U.
Secondly, several experimental observations suggest an
unconventional superconductivity in this material. Both of these
important aspects are the subjects of our investigation in which Pr
is systematically replaced by La.

In our previous study\cite{Rotundu} we have demonstrated the usefulness of this
alloying method due to the fact that both end-compounds
PrOs$_{4}$Sb$_{12}$ and LaOs$_{4}$Sb$_{12}$ exist in the same
crystal structure with essentially identical lattice constants. This
lack of a mismatch of the lattice parameters provides a unique
opportunity to study the superconducting state that is normally very
sensitive to local stresses resulting from a distribution of
interatomic distances. Also, the fact that both end-compounds are
superconducting with $T_c$'s differing only by a factor of 2
allows one to compare physical properties in closely related
superconductors of which one is conventional while the other is
probably not. Among experimental evidences suggesting the
unconventional character of superconductivity in PrOs$_{4}$Sb$_{12}$
are the temperature dependence of the specific heat\cite{Bauer},
spontaneous appearance of a static magnetic field bellow the
superconducting transition\cite{Aoki}, penetration depth
mesaurements\cite{Broun,Chia}, and presence of two superconducting
transitions in the specific 
heat\cite{Maple,Vollmer,Aoki2,Measson,Cichorek,Rotundu} and thermal conductivity 
measurements\cite{Izawa}. However, neither of the
interpretations of these observations is well established
and some of them remain controversial. In particular, there is no
consensus whether the two transitions in the specific heat coexist
or whether they occur at different parts of the sample at different
temperatures.

Equally controversial is the origin of the heavy fermion behavior in
PrOs$_{4}$Sb$_{12}$. Theoretical scenarios most often considered
include the quadrupolar Kondo effect\cite{Cox}, inelastic scattering
by low-energy crystal field levels\cite{Goremychkin,Fulde}, rattling 
motion\cite{Goto} of rare-earth ions, and fluctuations of the
quadrupolar order parameter. However, a conventional superconductivity and lack 
of mass enhancement in LaOs$_{4}$Sb$_{12}$ indicate an importance of $f$-
electrons. The first two models on the other hand, being single
impurity type models, seem to be inconsistent with the results of the
specific heat study of Pr$_{1-x}$La$_x$Os$_{4}$Sb$_{12}$. The $C/T$ (specific 
heat
divided by temperature) discontinuity at $T_c$, which is
proportional to $m^*$, is strongly suppressed by the La doping\cite{Rotundu}. At
the same time, the relevant single-impurity parameters, such as the
crystalline electric field (CEF) spectrum and interatomic separations, are 
essentially unaltered by the
alloying. However, since the $C/T$ discontinuity at $T_c$ depends
also on a coupling strength, we have searched for additional
evidences of the collective nature of the heavy fermion state in
PrOs$_{4}$Sb$_{12}$. Here we report the results for the upper critical
field of Pr$_{1-x}$La$_{x}$Os$_{4}$Sb$_{12}$ studied by specific heat.

\section{Experimental}

Samples used in this investigation were single crystals grown by
the Sb-flux method. All specific heat measurements were performed on individual 
single crystals. These crystals were between 1 and 5 mg. 

Before discussing results for the critical fields we comment on the ac-
susceptibility data shown in Fig. 1. The ac-susceptibility measurements were 
performed in the Earth's magnetic field and ac field of approximately 0.1 Oe. No 
significant differences in these results were detected for two different 
frequencies used, 27 and 273 Hz. The ac-susceptibility, arbitrarily normalized, 
is shown versus reduced temperature, $T/T_c$, where $T_c$ for this purpose was 
determined by the diamagnetic onset. The results of the ac-susceptibility are 
sample dependent, similarly to the discussed next specific heat. For $x=0$ we 
show the data for two crystals, of which the crystal No. 2 clearly has two steps 
in the ac-susceptibility, while crystal No. 1 seems to have one step only. 
However, both samples have wide transitions. The transition for the crystal with 
a single step is even wider than the combined transition of crystal No. 2. All 
previously reported ac-susceptibility curves for $x=0$ also showed large 
widths\cite{Measson, Cichorek, Drobnik}, typically 0.2 K. The temperature 
separation of the two steps in the susceptibility, 0.13-0.14 K, is approximately 
equal to the temperature separation of the two peaks in the specific heat. These 
results strongly argue for the inhomogeneous coexistence of two superconducting 
phases in PrOs$_{4}$Sb$_{12}$. Substitution of small amounts of La for Pr (5\%) 
does not eliminate the two steps in the susceptibility but makes the separation 
of the two steps and the overall width of the transition smaller. In general, 
the ac-susceptibilities for samples in the range $x=0.2$ to 0.8 are similar. 
There is a slight increase of the width of the transition (on the reduced 
temperature scale) between $x=0.6$ and 0.8, followed by a dramatic reduction for 
LaOs$_{4}$Sb$_{12}$. The reduction of the width of the transition in the ac-
susceptibility with small values of $x$ correlates with our previous observation 
of the decrease of the width of the superconducting anomaly in the specific heat 
between $x=0$ and a crossover concentration $x_{cr} \approx 0.2-
0.3$.\cite{Rotundu} The specific heat implied the disappearance of the second 
superconducting transition near $x_{cr}$.

These wide transitions in the ac-susceptibility of PrOs$_4$Sb$_{12}$ cannot be 
explained by very small $H_{c1}$ (field at which magnetic flux is completely 
expelled) in comparison with $H_{c2}$ (onset of diamagnetism). Based on the 
reported\cite{Cichorek} $dH_{c1}/dT$ slope we expect the width of the transition 
to be less than 20 mK. The origin of possible inhomogeneities is unclear. Sharp 
transitions observed in LaOs$_{4}$Sb$_{12}$ samples argue against structural 
defects since La and Pr are chemically similar. Thus, these inhomogeneities seem 
to be associated with $4f$-electrons of Pr. One plausible scenario is a mixture 
of two electronic configurations of Pr, $4f^1$ and $4f^2$. The analysis of the 
high temperature magnetic susceptibility by various groups lead to the effective 
high temperature paramagnetic moment between 3.2 and 3.6 $\mu$$_B$/Pr. (The 
expected moments are 2.54 and 3.59 $\mu$$_B$ for $4f^1$ and $4f^2$ 
configurations, respectively.) Small size of crystals and sample holder 
contribution could lead to these discrepancies. Therefore, we have synthesized a 
large 50 mg crystal, which was placed between two long concentric and 
homogeneous tubes such that no background subtraction was needed to accurately 
determine the magnetization due to the crystal. The effective magnetic moment 
measured for this crystal at temperatures between 200 and 350 K is equal to the 
expected value for the $4f^2$ configuration. Also, the $L_{III}$ 
absorption\cite{MapleL3} and inelastic neutron 
scattering\cite{Goremychkin,Kuwahara} results argue for the average Pr valence 
very close to 3. Another possible scenario for the existence of inhomogeneities 
is due to the closeness of the system to a long range antiferro-quadrupolar 
order.\cite{Aoki2} Clusters with a short-range-order would have different 
superconducting parameters than the remaining part of the sample. The discussed 
next magnetic response of the specific heat shows that these two plausible 
superconducting phases might not be independent; there seems to be some coupling 
between them. 

There is also some sample dependence of the two superconducting anomalies in the 
specific heat of the pure PrOs$_{4}$Sb$_{12}$ and of their response to small 
magnetic fields. Let us consider the zero field data first. Crystal No. 1, for 
which a wide transition in the ac-susceptibility is shown in Fig.1, (most upper 
panel) has rather well defined anomalies in $C/T$ at temperatures $T_{c1}$ and 
$T_{c2}$. (The method of extracting $T_{c1}$ and $T_{c2}$ for the purpose of 
critical fields' analysis is illustrated in Fig. 2.) Another investigated 
crystal from the same batch exhibited almost identical zero-field specific heat. 
On the other hand crystal No. 2 from a different batch, that showed two steps in 
the ac-susceptibility, has a sharp peak at $T_{c2}$ and only a small shoulder 
that might correspond to a transition at $T_{c1}$. This strong sample dependence 
can be inferred from previously published specific heat data, as well. However, 
according to the majority of published specific heat measurements, the anomaly 
at $T_{c2}$ is more pronounced than that at $T_{c1}$.\cite{Vollmer,Frederick}  

We have also observed different response of the two PrOs$_{4}$Sb$_{12}$ crystals 
to small magnetic fields. Magnetic fields were applied approximately in the 
(100) direction for all crystals investigated. For crystal No.1 magnetic field 
suppresses the height of the anomaly at $T_{c2}$ stronger
than at $T_{c1}$, such that only the high temperature transition
at $T_{c1}$ can be clearly seen in fields of the order 0.5 T. In
particular, the negative slope in $C/T$ versus $T$ between $T_{c2}$
and $T_{c1}$ in $H=0$ T becomes positive in the field of 0.5 T. On the other 
hand, in sample No. 2 the low temperature anomaly dominates at all fields 
studied up to 1.2 T. This last crystal has both zero and magnetic field specific 
heat similar to those presented by Frederick et al.\cite{Frederick}

Comparisons of the susceptibility and zero-field specific heat for the two 
crystals seems to indicate different volume fractions occupied by two 
superconducting phases, with crystal No. 1 having a relatively large fraction 
corresponding to the higher $T_c$ phase. However, the redistribution of the 
relative heights of the anomalies in crystal No. 1 by magnetic fields suggests 
some interdependence of the two superconducting phases. 

Insets to Fig. 2 present the critical fields versus $T$ determined by the 
specific heat. For sample no. 2 we were not able to determine $T_{c1}$ and only 
one line of transitions, that for $T_{c2}$, is shown. In agreement with previous 
reports we find a positive curvature in $H_{c2}$ versus $T$ (near $T_c$) in 
sample no. 1. This positive curvature in $H_{c2}(T)$ for PrOs$_{4}$Sb$_{12}$ has 
been detected in various measurements, such as electrical resistivity, ac-heat 
capacity, magnetic susceptibility, and has been proposed to be a signature of 
the two-band superconductivity. Additional measurements of a part of this 
crystal showed that $H_{c2}$ between 0.3 and at least 0.6 T is linear in $T$. On 
the other hand, we do not find any curvature in $H_{c2}$ (extracted from the 
measurement of $T_{c2}$ versus $T$) near $T_{c2}$ in sample no. 2, within the 
resolution of this measurement.  

A small, but detectable curvature, in $H_{c2}$ versus $T$ was observed again in 
some Pr$_{1-x}$La$_{x}$Os$_{4}$Sb$_{12}$ crystals with small amounts of 
Lanthanum, $x=0.02$ and 0.05. Figure 3 shows $C/T$ for one of our $x=0.05$
alloys in several representative fields. (Only one line of transitions could be 
clearly identified for all alloys with $x>0$.) There might be some curvature in 
the $H_{c2}(T)$ for $x=0.1$ (Fig. 4), although our data are not sufficiently 
precise to resolve it. On the other hand, alloy $x=0.3$ (Fig. 5) and alloys with 
$x>0.3$ seem to have a conventional variation of $H_{c2}(T)$, without a positive 
curvature near $T_c$. 

The slope of $H_{c2}$ versus $T$ at $T_{c}$ is of great interest
since it is related to the effective mass of Cooper pairs. The
discussed plausible curvature makes the determination of
$dH_{c2}/dT$ for the pure Pr compound and weakly doped alloys
somewhat arbitrary. Furthermore, there is an important question
whether $H_{c2}$ is linear in $T$ in the direct vicinity of $T_{c1}$
followed by a crossover regime to a linear dependence with a larger slope at 
lower temperatures or if there is a curvature at any temperature near $T_{c1}$. 
Our point by point specific heat measurement technique with limited resolution 
can not fully address this issue but seems to prefer the first scenario. For 
several small $x$-value alloys we were able to observe two linear regimes. 
Microwave surface impedance measurements by Broun et al.\cite{Broun} clearly 
identify two linear regimes in $\lambda^{-2}$ versus $T$ near $T_c$. However, 
the relationship between our critical field measurement and reported penetration 
depth is not clear at present. 

Therefore, in Fig. 5 we show -$dH_{c2}/dT$ calculated for sufficiently large 
fields, for which $H_{c2}$ is clearly linear in $T$ (for $H>2000$ Oe for $x=0$ 
and $H>1000$ for $x=0.02$ and 0.05), as closed symbols and the initial slope 
measured in fields smaller than 1000 Oe
as open symbols. In agreement with previous studies this critical
field slope for PrOs$_{4}$Sb$_{12}$, obtained in this case for fields larger 
than 0.5 kOe, is about -2.1T/K. -$dH_{c2}/dT$ decreases monotonically with $x$, 
but
most of this reduction takes places for small values of $x$. Such a
variation of the slope strongly implies that the effective mass of
carriers is rapidly reduced by a small amount of La substituted for
Pr.

As it has been already demonstrated, the superconductivity of
PrOs$_{4}$Sb$_{12}$ can be considered in the clean limit. The value
of the mean free path $l_e$ is between 1000 and 2000 {\AA}. This lower limit is 
obtained from our resistivity measurement ($\rho_0$ approximately 5 $\mu\Omega$-
cm) while the upper limit from the Dingle temperature\cite{Aoki,Dingle}.  Thus, 
the mean free path is significantly larger than the
reported coherence length $\xi_0$ of 116 {\AA} calculated from the critical
field slope. Our best estimate of $\xi_0$, 112 {\AA}, is in excellent agreement 
with that value, considering uncertainties of the measurements. We expect the 
mean free path for LaOs$_{4}$Sb$_{12}$ to be of the same order as that for 
PrOs$_{4}$Sb$_{12}$. Using the published $\rho_0$ of approximately 3 
$\mu\Omega$-cm we arrive at $l\approx 1600$ {\AA}. The coherence length for 
LaOs$_{4}$Sb$_{12}$ is much larger than that for PrOs$_{4}$Sb$_{12}$. Assuming 
the validity of the clean limit formulae, $\xi_0 \approx$ 970 {\AA}, in 
agreement with the approximation of Sugawara et al.\cite{Sugawara} The 
corresponding Fermi velocities are 1.52 10$^6$ and 5.25 10$^6$ cm/s for the Pr 
and La compound, respectively. Since the lattice constants of both compounds are 
almost identical and Pr has the same number of valence electrons as La, the 
Fermi wave vectors should be equal. Therefore, the ratio of Fermi velocities 
should be equal to the ratio of effective masses or electronic specific heat 
coefficients. There is some distribution for reported $\gamma$ values of 
LaOs$_{4}$Sb$_{12}$. Using the largest reported value\cite{Rotundu} of 59 
mJ/K$^2$ and the derived ratio of Fermi velocities we arrive at 200 mJ/K$^2$ as 
the estimate of $\gamma$ for PrOs$_{4}$Sb$_{12}$. This estimate is in contrast 
to the discontinuity in $\Delta C/T_c/\gamma$, which falls between 800 - 1000 
mJ/K$^2$ mol (depending on sample and analysis method), suggesting an 
enhancement of $\Delta C/T_c/\gamma$ due to collective many-body effects beyond 
mass enhancement.

\begin{figure}
%h=here, t=top, b=bottom, p=separate figure page
\begin{center}
\leavevmode
\includegraphics[width=0.84\linewidth]{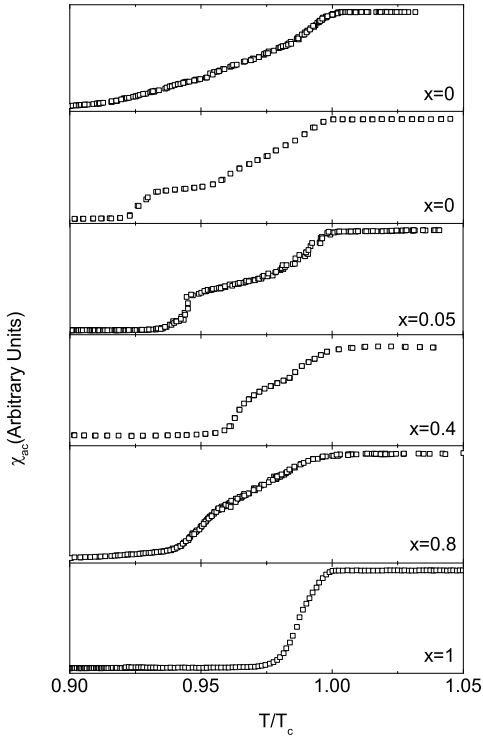}
\caption {ac-susceptibility versus reduced temperature for representative 
crystals.}\label{fig1}\end{center}\end{figure}

\begin{figure}
%h=here, t=top, b=bottom, p=separate figure page
\begin{center}
\leavevmode
\includegraphics[width=0.84\linewidth]{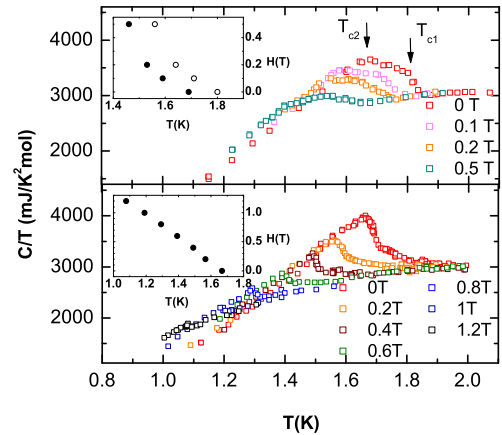}
\caption {$C/T$ versus $T$, near $T_c$, for two PrOs$_4$Sb$_{12}$ samples (No.1 
in the upper panel; No. 2 in the lower panel) in magnetic
fields. The inset shows $H_{c2}$ versus $T$. See on-line for 
color.}\label{fig2}\end{center}\end{figure}

\begin{figure}
%h=here, t=top, b=bottom, p=separate figure page
\begin{center}
\leavevmode
\includegraphics[width=0.84\linewidth]{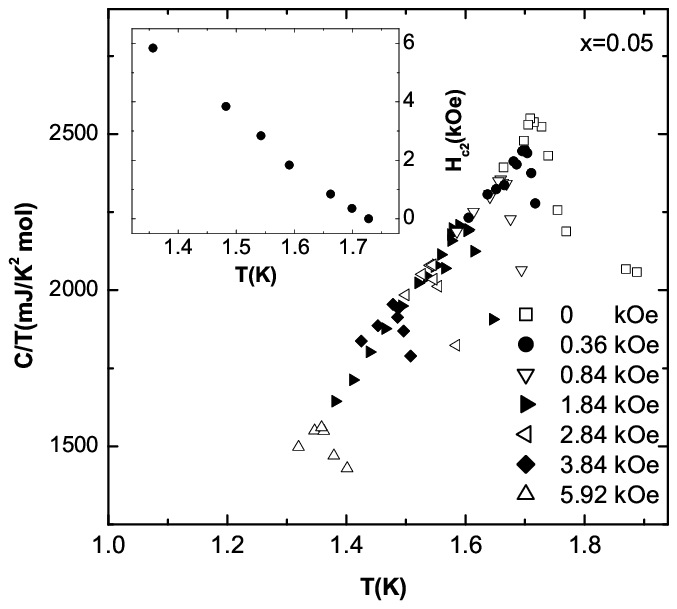}
\caption {$C/T$ versus $T$, near $T_c$, for $x=0.05$ in magnetic
 fields. The inset shows $H_{c2}$ versus 
$T$.}\label{fig3}\end{center}\end{figure}

\begin{figure}
%h=here, t=top, b=bottom, p=separate figure page
\begin{center}
\leavevmode
\includegraphics[width=0.84\linewidth]{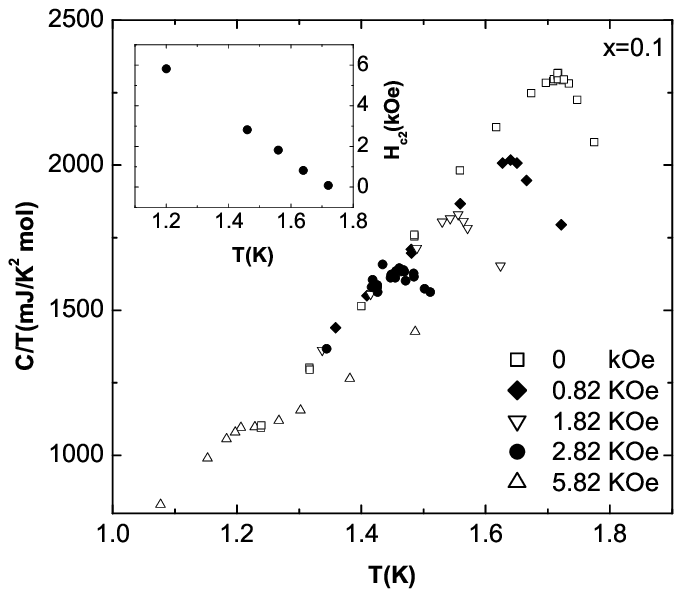}
\caption {$C/T$ versus $T$, near $T_c$, for $x=0.1$ in small magnetic
 fields. The inset shows $H_{c2}$ versus 
$T$.}\label{fig4}\end{center}\end{figure}

In the clean-limit of superconductivity, which appears to be the appropriate 
limit for Pr$_{1-x}$La$_{x}$Os$_{4}$Sb$_{12}$ alloys, the
effective mass depends on $\sqrt{-dH_{c2}/dT/T_c}$. Figure 7
shows this latter quantity versus concentration $x$. Again, this figure
suggests that the greatest reduction of the effective mass takes
place for small values of $x (<0.2-0.3)$. Interestingly, when
$\sqrt{-dH_{c2}/dT/T_c}$ is extrapolated from high and
intermediate values of $x$ to $x=0$, the obtained value is close to the
one calculated from the initial slope of $dH_{c2}/dT$ for PrOs$_{4}$Sb$_{12}$. 
This
observation provides an indirect support to our hypothesis that there are two 
linear
regimes in $H_{c2}$ versus $T$ in at least some crystals of the undoped Pr 
compound.

\begin{figure}
%h=here, t=top, b=bottom, p=separate figure page
\begin{center}
\leavevmode
\includegraphics[width=0.84\linewidth]{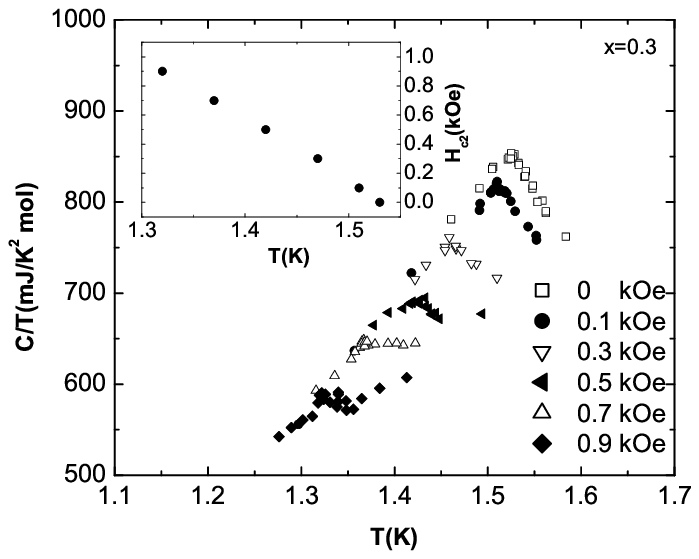}
\caption {$C/T$ versus $T$, near $T_c$, for $x=0.3$ in magnetic
 fields. The inset shows $H_{c2}$ versus 
$T$.}\label{fig5}\end{center}\end{figure}

\begin{figure}
%h=here, t=top, b=bottom, p=separate figure page
\begin{center}
\leavevmode
\includegraphics[width=0.84\linewidth]{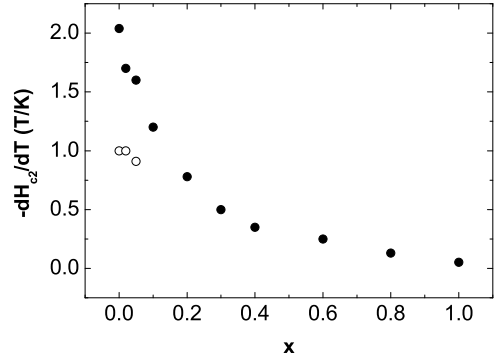}
\caption {-$dH_{c2}/dT$ versus $x$. Open circles for $x<0.1$ correspond to -
$dH_{c2}/dT$ measured directly at $T_c$. See text for 
details.}\label{fig6}\end{center}\end{figure}

\begin{figure}
%h=here, t=top, b=bottom, p=separate figure page
\begin{center}
\leavevmode
\includegraphics[width=0.84\linewidth]{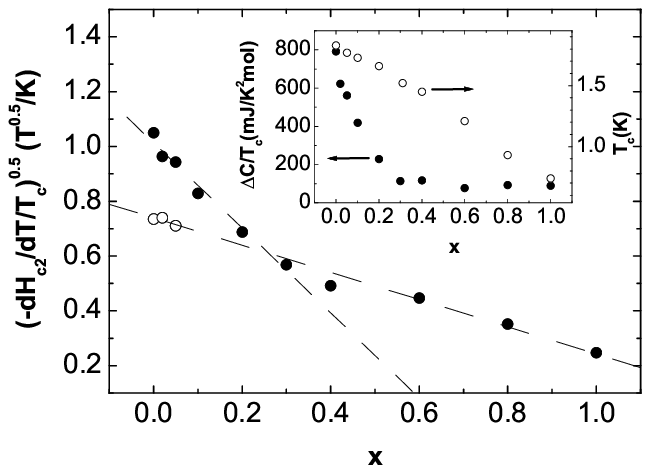}
\caption {$\sqrt(-dH_{c2}/dT/T_c)$ versus $x$. Open circles denote $\sqrt(-
dH_{c2}/dT/T_c)$ at $T_c$ (see text for details). The dashed lines are guides to 
the eye. The inset represents the discontinuity of $C/T$ at $T_c$ and $T_c$ 
(identified by the maxima\cite{Rotundu} in $C/T$) versus 
$x$.}\label{fig7}\end{center}\end{figure}

In the inset to Fig. 7 we present the overall discontinuity of the
specific heat divided by $T_{c}$. This latter quantity can
also be used as an estimate of the effective electron mass,
$m^*$, particularly if the coupling strength were constant. Note a close 
resemblance of the concentration dependence of
the two sets of data, $\sqrt{-dH_{c2}/dT/T_c}$ and
$\Delta C/T_c$. Clearly, $x_{cr} \approx 0.2-0.3$ is a crossover
concentration between a rapid reduction of these quantities upon
the La-doping and regime corresponding to moderate or small changes with $x$.
The critical field slope (Fig. 5 and 6) implies that there is still
significant reduction of $m^*$ with $x$ for $x>0.3$, but the main change
takes place for $x<x_{cr}$. A crossover concentration for $m^*$ seems to
exist also in Pr(Os$_{1-x}$Ru$_{x}$)$_{4}$Sb$_{12}$ series.\cite{Frederick} Its
x$_{cr}$ is much smaller, of about 0.06 only. However, the Ru-alloying is 
expected to have much more drastic effects on {f}-electrons of Pr than
the La-doping. The hybridization parameters are directly
altered by the Ru-alloying and the CEF energies are increased.

Consider this positive curvature or two slopes in $H_{c2}$ versus $T$ of
PrOs$_{4}$Sb$_{12}$ in the framework of the two-band superconductivity
model. A smaller initial slope seems to suggest that the transition
at $T_{c1}$ is dominated by lighter quasiparticles. Heavier
electrons contribute mainly to the lower temperature transition.
However, our results suggest that the ratio of the effective masses
of the two bands is of the order 2, only. This could explain why the
two discontinuities in $C(T)$ at $T_{c1}$ and $T_{c2}$ in some of the reported 
results are of similar magnitude. Furthermore, the La-alloying data indicate 
that the
change of slope in $H_{c2}$ versus $T$, near $T_c$, correlates with the 
existence
of two superconducting transitions in the specific heat. I.e., alloys with 
$x<x_{cr}$ in addition to showing enhanced $\Delta C/T_c$, usually exhibit a 
curvature or two different slopes in $H_{c2}$ versus $x$ and wide or two 
transitions in the specific heat. This correlation between the existence of the 
curvature in $H_{c2}$ versus $x$, argued to be due to two-band 
superconductivity, and presence of two superconducting transitions is 
unexpected.  

\section{Summary and Acknowledgements}

The change of the slope in $H_{c2}$ versus $T$ near $T_c$, previously ascribed 
to two-band superconductivity,\cite{Measson} seems to be characteristic of 
samples and concentrations that exhibit two superconducting anomalies. This 
correlation is clearly inconsistent with the proposed Josephsone coupling of the 
two bands.\cite{Broun, Measson} In this latter case, only the higher temperature 
transition should be observed. On the other hand, if we consider the No. 1 
crystal of PrOs$_4$Sb$_{12}$ only, for which we clearly observe the curvature in 
$H_{c2}$ versus $T$, the magnetic field response could be understood if the 
coupling between the two order parameters was brought about by external magnetic 
fields. 

The reported temperature variation of the upper
critical field in Pr$_{1-x}$La$_{x}$Os$_{4}$Sb$_{12}$, obtained from
specific heat measurements, implies the existence of a crossover concentration,
$x_{cr}\sim 0.25$, at which both the rate at which -$dH_{c2}/dT$ varies with $x$ 
and the character of the temperature variation of $H_{c2}$ change. The same 
crossover
concentration was previously identified in $\Delta C/T_c$. The
positive curvature (or change in slope) in $H_{c2}$ versus $T$
persists in some crystals with small values of $x$, possibly to $x_{cr}$. 

Finally, our results suggest strong sample dependence of superconducting 
properties of PrOs$_4$Sb$_{12}$. The ac-susceptibility seems to indicate 
the presence of inhomegeneities that are difficult to account for considering sharp 
transitions in the isostructural and chemically almost identical 
LaOs$_4$Sb$_{12}$. On the other hand, assuming homogeneous samples, the upper 
superconducting transition in sample No. 2 might be related to the phase 
transition with an order higher than 2.\cite{Kumar} In such a case, strong 
sample dependence and ac-susceptibility mimicking a two-step transition can be 
expected. A higher order phase transition (such as the third order) would be 
very susceptible to impurities and imperfections leading to the 2$^{nd}$ order 
transition in sufficiently imperfect crystals. The expulsion of magnetic flux 
would be weak upon lowering temperature from $T_{c1}$ to $T_{c2}$ followed by a 
much more rapid expulsion below the lower superconducting transition. Also, this 
scenario could account for different response to magnetic field of two crystals 
exhibiting different order phase transitions at $T_{c1}$. However, the 
possibility of the 3$^{rd}$ order phase transition remains speculative since few 
materials exhibiting phase transitions with an order higher than 2 were 
reported.\cite{Kumar}

This work was supported by the Department of Energy, grant No. DE-FG02-
99ER45748, National Science Foundation, grant No. DMR-0104240, and Grant-in-Aid 
for Scientific Research Priority Area "Skutterudite" (No. 
15072206) of the Ministry of Education, Culture, Sports, Science and  
Technology, Japan.


\begin{thebibliography}{12}

\bibitem{Bauer}
E.D. Bauer, N.A. Frederick, P.-C. Ho, V.S. Zapf, and M.B. Maple,  Phys.\ Rev.\ B 
{\bf 65}, 100506(R)(2002).

\bibitem{Rotundu}
C.\ R.\ Rotundu, P. Kumar, and B. Andraka, Phys.\ Rev.\ B {\bf 73}, 014515 
(2006).

\bibitem{Aoki}
Y.\ Aoki, T. Tsuchiya, T.\ Kanayama, S.R. Saha, H. Sugawara, H. Sato, W. 
Higemoto, A. Koda, K. Ohishi, K. Nishiyama, and R. Kadono, Phys.\ Rev.\ Lett.\ 
{\bf 91}, 067003 (2003).

\bibitem{Broun}
D.M. Broun, P.J. Turner, G.K. Mullins, D.E. Sheehy, X.G. Zheng, S.K. Kim, N.A. 
Frederick, M.B. Maple, W.N. Hardy, and D.A. Bonn, cond-mat/0310613.

\bibitem{Chia}
E. E. M. Chia, M. B. Salamon, H. Sato, and H. Sugawara, Phys.\ Rev.\ B {\bf 69}, 
180509 (2004).

\bibitem{Maple}
M.\ B.\ Maple et al., J.\ Phys.\ Soc.\ Jpn.\ {\bf 71}, Suppl.\ 23 (2002).

\bibitem{Vollmer}
R.\ Vollmer, A. Faißt, C. Pfleiderer, H. v. Löhneysen, E. D. Bauer, P.-C. Ho, V. 
Zapf, and M. B. Maple, Phys.\ Rev.\ Lett.\ {\bf 90}, 057001 (2003).

\bibitem{Aoki2}
Y.\ Aoki, T. Namiki, S. Ohsaki, S.R. Saha, H. Sugawara, and H. Sato, J.\ Phys.\ 
Soc.\ Jpn.\ {\bf 71}, 2098 (2002).

\bibitem{Cichorek}
T.\ Cichorek, A.C. Mota, F. Steglich, N.A. Frederick, W.M. Yuhasz, and M.B. 
Maple, Phys.\ Rev.\ Lett.\ {\bf 94}, 107002 (2005).

\bibitem{Measson}
M.-A. Measson, D. Braithwaite, J. Flouquet, G. Seyfarth, J.P. Brison, E. Lhotel, 
C. Paulsen, H. Sugawara, and H. Sato, Phys.\ Rev.\ B {\bf 70}, 064516 (2004).

\bibitem{Izawa}
K. Izawa, Y. Nakajima, J. Goryo, Y. Matsuda, S. Osaki, H. Sugawara, H. Sato, P. 
Thalmeier, and K. Maki, Phys.\ Rev.\ Lett.\ {\bf 90}, 117001 (2003).

\bibitem{Cox}
D.\ Cox, Phys.\ Rev.\ Lett.\ {\bf 59}, 1240 (1987).

\bibitem{Goremychkin}
E.\ A.\ Goremychkin, R. Osborn, E.D. Bauer, B. Maple, N.A. Frederick, W.M. 
Yuhasz, F. M. Woodward, and J.W. Lynn, Phys.\ Rev.\ Lett.\ {\bf 93}, 157003 
(2004).

\bibitem{Fulde}
P.\ Fulde and J. Jensen, Phys.\ Rev.\ B {\bf 27}, 4085 (1983).

\bibitem{Goto}
T. Goto, Y. Nemoto, K. Sakai, T. Yamaguchi, M. Akatsu, T. Yanagisawa, H. Hazama, 
and K. Onuki, Phys.\ Rev.\ B {\bf 69}, 180511 (2004).

\bibitem{Drobnik}
S. Drobnik, K. Grube, C. Pfleiderer, H. v. Löhneysen, E. D. Bauer, and M.B. 
Maple, Physica B {\bf 359-361}, 901 (2005).

\bibitem{MapleL3}
D. Cao, F. Bridges, M.B. Maple, E.D. Bauer, and S. Bushart, Phys.\ Rev.\ B {\bf 
67}, 180511 (2003).

\bibitem{Kuwahara}
K. Kuwahara, K. Iwasa, M. Kohgi, K. Kaneko, N. Metoki, S. Raymond, M.-A. 
Measson, J. Flouquet, H. Sugawara, Y. Aoki, and H. Sato, Phys.\ Rev.\ Lett.\ 
{\bf 95}, 107003 (2005).

\bibitem{Frederick}
N.A. Frederick, T.A. Sayles, S.K. Kim, and M.B. Maple, cond-mat/0511130 (2005).

\bibitem{Dingle}
H. Sugawara, S. Osaki, S.R. Saha, Y. Aoki, H. Sato, Y. Inada, H. Shishido, R. 
Settai, Y. Onuki, H. Harima, and K. Oikawa, Phys.\ Rev.\ B {\bf 66}, 220504(R) 
(2002).
 
\bibitem{Sugawara}
H. Sugawara, M. Kobayashi, S. Osaki, S.R. Saha, T. Namiki, Y. Aoki, and H. Sato, 
Phys.\ Rev.\ B {\bf 72}, 014519 (2005).

\bibitem{Kumar}
P. Kumar, D. Hall and R. G. Goodrich, Phys.\ Rev.\ Lett.\ {\bf 82}, 4532 (1999); 
P. Kumar, Phys.\ Rev.\ B {\bf 68}, 064505 (2003); R. Werner and V. J. Emery, 
Phys.\ Rev.\ B {\bf 67}, 014504 (2003).

\end{thebibliography}
\end{document}